\documentclass[12pt]{article}

\usepackage{graphicx}
\begin{document}
\begin{center}
{\Large The Graviton as a Bound State and the Cosmological Constant
Problem}
\vskip 0.3 true in {\large J. W. Moffat}
\date{} \vskip 0.3 true in
{\it Perimeter Institute for Theoretical Physics,
Waterloo, Ontario N2J2W9, Canada}
\vskip 0.1 true in
and
\vskip 0.1 true in
{\it Department of Physics, University of Toronto, Toronto, Ontario
M5S1A7, Canada}

\end{center}

\begin{abstract}%
The graviton is pictured as a bound state of a fermion and anti-fermion
with the spacetime metric assumed to be a composite object of spinor fields, based on a
globally Lorentz invariant action proposed by Hebecker and
Wetterich. The additional degrees of freedom beyond those of
the graviton are described by Goldstone boson gravitational
degrees of freedom. If we assume that the fermion is a light
neutrino with mass $m_\nu\sim 10^{-3}$ eV, then we obtain the
effective vacuum density ${\bar\rho}_\lambda\sim (10^{-3}\,
eV)^4$, which agrees with the estimates for the cosmological
constant from WMAP and SNIa data.

\end{abstract}
\vskip 0.2 true in e-mail: john.moffat@utoronto.ca

\vskip 0.3 true in

\section{\bf Introduction}

The cosmological constant problem is generally accepted to be one of
the most serious paradoxes in modern particle physics and
cosmology~\cite{Weinberg,Straumann}. Let us express the cosmological
constant in the form
\begin{equation}
\lambda=\lambda_0+8\pi G\rho_{\rm vac},
\end{equation}
where $\lambda_0$ is the bare cosmological constant in the
Einstein-Hilbert action, and $\rho_{\rm vac}$ is the vacuum
contribution from the vacuum expectation value
\begin{equation}
\langle T_{\mu\nu}\rangle=g_{\mu\nu}\rho_{\rm vac}+{\rm higher\,
curvature\,terms}.
\end{equation}
We know from cosmological observations that
\begin{equation}
\rho_\lambda \sim\rho_{\rm crit}=\frac{3H_0^2}{8\pi G}=8\times
10^{-47}\,h_0^2\,GeV^4,
\end{equation}
where $\rho_\lambda=\lambda/8\pi G$ and $h_0=H_0/(100\,{\rm
km}\,s^{-1}\,{\rm Mpc}^{-1})\sim 0.6$.

From standard field
theory calculations, we find that calculations of $\rho_\lambda$ differ
from the value $\rho_\lambda\sim 10^{-47}\, GeV^4\sim (10^{-3}\, eV)^4$,
obtained from WMAP and SNIa data~\cite{Bennett,Spergel,Perlmutter,Riess},
by $10^{55}$ at the electroweak energy scale$\sim 250$ Gev, and by
$10^{121}$ in natural reduced Planck energy units.

The significant discrepancy between the expectations of particle physics
and the cosmological data strongly suggests that we cannot picture the
graviton as an ``ordinary'' particle, such as the photon and the $W$ and
$Z$ mesons of the standard model~\cite{Zee}. The shift in the action
$S\rightarrow S-S_{\rm vac}$ is made non-trivial by the fact that the
graviton couples to all forms of matter universally, whereas the photon
only sees electric charge and the gluon only sees color charge. Thus, the
solution of the cosmological constant problem must reside in a radical
change in how we picture the graviton. Various possible explanations have
been proposed for how this change should be
implemented~\cite{Moffat,Sundrum}. The cosmological constant problem has
been exacerbated by the problem of dark energy. If the vacuum energy
density is the source of dark energy, as is suggested by the WMAP data
analysis, then we no longer seek to find an explanation for $\lambda$
being zero, but we must explain the incredible smallness of $\lambda$ when
fitted to the cosmological data.

In the following, we
shall investigate the consequences of picturing the graviton as a bound
state condensate of a fermion and an anti-fermion. It has been known for a
long time that by postulating something akin to Sakharov's induced
gravity~\cite{Sakharov}, generically a one loop action
produces an effective action containing the Einstein-Hilbert action with a
cosmological constant term~\cite{Visser}. We will base
our desciption of the graviton condensate on a recently proposed
model of composite gravitons in which the metric tensor of
spacetime is described by the expectation value of a vierbein
spinor bilinear form~\cite{Hebecker,Wetterich}. By identifying the fermion field
as a light neutrino with mass $m_\nu\sim 10^{-3}$ eV, we predict that the
effective vacuum density ${\bar\rho}_\lambda\sim (10^{-3}\,
eV)^4$ in agreement with estimates from the WMAP and SNIa
data~\cite{Bennett,Spergel,Perlmutter,Riess}.

\section{\bf Bound State Model}

Let us assume that the graviton is a bound state of a fermion
particle. The metric is given by
\begin{equation}
\label{metric}
g_{\mu\nu}=\langle E^a_\mu
E^b_\nu\eta_{ab}\rangle,
\end{equation}
where $M^*_{PL}=M_{PL}/\sqrt{8\pi G}=2.4\times
10^{18}$ GeV is the reduced Planck mass and~\cite{Hebecker,Wetterich}:
\begin{equation}
\label{vierbein}
E^a_\mu=\frac{i}{2}\biggl(\frac{1}{M^*_{PL}}\biggr){\bar\psi}\gamma^a\partial_\mu\psi+h.c.
\end{equation}
We assume that $E={\rm det}(E^a_\mu)\not= 0$, so that $E^a_\mu$ and
$g^{\mu\nu}$ are well defined. We have scaled $\psi$ so that it,
$E^a_\mu$ and $g_{\mu\nu}$ have mass dimension 0.

A diffeomorphism and {\it globally} Lorentz invariant action
is~\cite{Hebecker,Wetterich}
\begin{equation}
\label{action}
S=M_{PL}^{*4}\int d^4xE(x),
\end{equation}
where
\begin{equation}
\label{Edeterminant}
E=\frac{1}{4!}\epsilon^{\mu_1...\mu_4}\epsilon_{a_1...a_4}E^{a_1}_{\mu_1}...
E^{a_4}_{\mu_4}=\biggl(\frac{1}{M^{*4}_{PL}}\biggr){\cal O},
\end{equation}
and ${\cal O}$ is a bilinear spinor operator.

In contrast to Einstein gravity, $S$ is not locally Lorentz
invariant, because the covariant derivative $D_\mu$ contains the
Christoffel connection ${\Gamma^\lambda}_{\mu\nu}$ but not a spin
connection. This leads to a generalized gravity theory in which the
vierbein contains additional degrees of freedom that are not described by
$g_{\mu\nu}$.  These additional degrees of freedom produce new invariants
not present in Einstein gravity, which are globally Lorentz invariant but
not locally Lorentz invariant. A nonlinear field decomposition
$E^a_\mu=e^b_\mu H^a_b$ can be employed, where $e^b_\mu$ describes the
standard Einstein gravity vierbein and $H^a_b$ describes the new
degrees of freedom, associated with Goldstone boson excitations due to
the spontaneous breaking of a global symmetry. The $H^a_b$ would
correspond in Einstein gravity to gauge degrees of freedom of the local
Lorentz transformation group and therefore not be present in an invariant
action. In the Hebecker-Wetterich (HW) action there exist new propagating
massless particles associated with the new degrees of freedom.

An estimate of the one loop order of fermionic fluctuations,
$\Gamma_{[E]}$, is given by~\cite{Hebecker,Wetterich}:
\begin{equation}
\label{fluctuations}
\Gamma_{[E]}=\alpha\int d^4x\langle E\rangle-\frac{1}{2}\ln(\langle
E\rangle{\cal D}),
\end{equation}
where $\alpha=(-1)^5M_{PL}^{*4}$
and $\langle E\rangle{\cal D}$ is the second functional derivative of the
bosonic action $S_B$ with respect to $\psi$. The bosonic action $S_B$ is
defined in terms of the partition function $Z$ as a functional integral
over the fermion field $\psi$ and a boson field $\chi$. Moreover, ${\cal
D}=\langle E^\mu_a\rangle\gamma^a{\hat D}_\mu$ where ${\hat D}_\mu=\partial_\mu
+(1/2\langle E\rangle)\langle E^a_\mu\rangle\partial_\nu(\langle EE^\nu_a\rangle)$.

The quantum field equations that follow from a variation of the effective
action $\Gamma(E)$ are given by
\begin{equation}
\frac{\delta\Gamma(E)}{\delta\langle E^a_\mu\rangle}=J^\mu_a,
\end{equation}
where $J^\mu_a=0$ in empty spacetime. An energy-momentum tensor which
includes all forms of matter and radiation is given by
\begin{equation}
\label{energymomentum}
T^{\mu\nu}=\langle E^{-1}E^{a\mu}\rangle J^\nu_a.
\end{equation}

The effective one loop action, obtained from the HW action
(\ref{action}), containing two derivatives, which is invariant under
global Lorentz transformations and diffeomorphism transformations
is
\begin{equation}
\label{effectiveaction}
\Gamma_{(2)}=\frac{M^{*2}_{PL}}{2}\int
d^4x\langle E\rangle[-R+\tau(D^\mu \langle E^\nu_a\rangle D_\mu
\langle E^a_\nu\rangle-2D^\mu \langle E^\nu_a\rangle D_\nu
\langle E^a_\mu\rangle)
$$ $$
+\beta D_\mu
\langle E^\mu_a\rangle D^\nu \langle E^a_\nu\rangle]+M^{*2}_{PL}\int
d^4x\langle E\rangle\lambda,  \end{equation}  where
\begin{equation}
D_\mu \langle E^a_\nu\rangle=\partial_\mu\langle E^a_\nu\rangle
-{\Gamma^\lambda}_{\mu\nu}\langle E^a_\lambda\rangle.
\end{equation}
Here, $\lambda$ is the cosmological constant in the
effective action, $\lambda=\rho_\lambda/M^{*2}_{PL}$,
${\Gamma^\lambda}_{\mu\nu}$ and $R$ are the  Christoffel symbols and Ricci
scalar obtained from the metric $g_{\mu\nu}$, and indices are raised and
lowered with the metric tensor. Due to the missing spin connection in
$D_\mu$, the two terms multiplying $\tau$ and $\beta$ are invariant only
under global Lorentz transformations.

\section{Resolution of the Cosmological Constant Problem}

According to Eq.(\ref{Edeterminant}), the cosmological constant
term $\sqrt{-g}\rho_\lambda$ in the Einstein-Hilbert action is
replaced in the effective action (\ref{effectiveaction}) by
\begin{equation}
{\bar\rho}_\lambda=\langle
E\rangle\rho_\lambda=\biggl(\frac{1}{M^{*4}_{PL}}
\biggr)\langle{\cal O}\rangle\rho_\lambda.
\end{equation}
We set the mass scale of $\langle{\cal O}\rangle$ by
\begin{equation}
\langle{\cal O}\rangle\sim O(m^4),
\end{equation}
where $m$ is a low energy mass.

We now have
\begin{equation}
{\bar\rho}_\lambda\sim
\biggl(\frac{m}{M^*_{PL}}\biggr)^4M^{*2}_{PL}\lambda.
\end{equation}
We choose $\lambda\sim M_{PL}^{*2}$ which is the result obtained for
a cutoff $\Lambda_c\sim M_{PL}^{*2}$ in natural reduced Planck units and we
obtain
\begin{equation}
{\bar\rho}_\lambda\sim m^4.
\end{equation}
Identifying the fermion field $\psi$ with the lightest neutrino field
$\psi_\nu$ with a mass $m=m_\nu\sim 10^{-3}$
eV~\cite{particledata,King,Giunti}, we get
\begin{equation}
\label{cc}
{\bar\rho}_\lambda\sim (10^{-3}\,eV)^4,
\end{equation}
which agrees with estimates for the vacuum density obtained from the WMAP
and SNIa data~\cite{Bennett,Spergel,Perlmutter,Riess}.

The terms multiplying $\tau$ and $\beta$ in (\ref{effectiveaction}) must be
small in order to avoid deviations from Einstein gravity and observations.
The contribution associated with $\beta$ vanishes in one loop order. An
analysis~\cite{Hebecker,Wetterich} of these contributions shows that for $\beta=0$,
the terms multiplying $\tau$ do not affect the lowest order post-Newtonian
gravity, and the Schwarzschild solution. For the
Friedmann-Robertson-Walker cosmological solution, a value of the Planck
mass is obtained which differs from Einstein gravity, which can be checked
by comparing it with a local gravitational measurement of Newton's constant
$G$.

The $\rho_{\rm vac}$ obtained from $T^{\mu\nu}_{\rm vac}$
in (\ref{energymomentum}) contains all contributions, including
those arising from phase transitions, such as chiral symmetry breaking, QCD
gluon condensates and Higgs spontaneous symmetry breaking in the standard
model.

A determination of the absolute values of neutrino masses is
difficult to achieve experimentally. Limits on neutrino masses
can be obtained from neutrino oscillation models, tritium decay
and cosmological bounds~\cite{King,Giunti}. From the mass
hierarchy of three-neutrino mixing models one finds in a normal
scheme the effective mass of the lightest neutrino has a value
between about $3\times 10^{-3}$ eV and $2\times 10^{-2}$ eV.
Hopefully, future experiments will narrow down the range of
values of the lightest neutrino mass.

\section{Conclusions}

By using a generalized gravity theory based on a bilinear operator form
developed by Hebecker and Wetterich~\cite{Hebecker,Wetterich}, we have
predicted a vacuum density ${\bar\rho}_\lambda\sim
(10^{-3}\,eV)^4$ in agreement with $\lambda CDM$ model estimates
from WMAP and SNIa data, when we identify the bound state fermion
associated with the graviton condensate with a light neutrino
with mass $m=m_\nu\sim 10^{-3}$ eV.

By identifying $\psi$ with a light neutrino field, we have predicted the
correct magnitude of ${\bar\rho}_\lambda$ that fits the $\lambda
CDM$ model interpretation of dark energy. This suggests that we
describe the dark energy as graviton condensates formed from
fluctuating light neutrinos. The source of dark energy would be
light neutrino and anti-neutrino condensates.

It would be interesting to investigate further whether there exists a
self-consistent gravity theory for composite gravitons, based on fermion
bilinear correlation functions that exhibits local Lorentz
invariance as opposed to the global Lorentz invariance of the HM
model. Up till now no such model has been discovered.

\vskip 0.3 true in

{\bf Acknowledgments}
\vskip 0.2 true in
This work was supported by the Natural Sciences and Engineering Research Council of
Canada.
\vskip 0.5 true in

\end{document}